\documentclass[sn-mathphys]{sn-jnl}

\jyear{2021}%

\theoremstyle{thmstyleone}%
\theoremstyle{thmstyletwo}%

\theoremstyle{thmstylethree}%

\begin{document}

\title[Antiferromagnetic Spin Pumping via Hyperfine Interaction]{Antiferromagnetic Spin Pumping via Hyperfine Interaction}


\author*{\fnm{Adam B.} \sur{Cahaya}}\email{adam@sci.ui.ac.id}

\affil*[1]{\orgdiv{Department of Physics}, \orgname{Universitas Indonesia}, \orgaddress{\street{Kampus UI Depok}, \city{Depok}, \postcode{16424}, \state{Jawa Barat}, \country{Indonesia}}}


\abstract{Spin pumping is an interfacial spin current generation from the ferromagnetic layer to the non-magnetic metal at its interface. The polarization of the pumped spin current $\textbf{J}_s \propto \textbf{m}\times \dot{\textbf{m}}$ depends on the dynamics of the magnetic moment $\textbf{m}$. When the materials are based on light transition metals, mechanism behind the spin current transfer is dominated by the exchange interaction between spin of localized d-electrons and itinerant conduction electrons. In heavier transition metals, however, the interaction is not limited to the exchange interaction. The spin of the conduction electron can interact to its nuclear spin by means of hyperfine interaction, as observed in the shift of NMR frequency. By studying the spin polarization of conduction electron of the non-magnetic metallic layer due to a nuclear magnetic moment $\textbf{I}$ of the ferromagnetic layer, we show that the hyperfine interaction can mediate the spin pumping. The polarization of the spin current generation is shown to have a similar form $J_s\propto \textbf{I}\times\dot{\textbf{I}}$.
}

\keywords{Dipole-dipole interaction, Hyperfine interaction, Spin current, Spin pumping, Spin mixing conductance}



\maketitle

\section{Introduction}\label{sec1}

Starting from the discovery of the giant magnetoresistance, the research area of spintronics has emerged. One of its focuses is the manipulation of magnetic moment by spin current and vice-versa \cite{PhysRevB.72.024426}. At magnetic interface, a spin current can be pumped from ferromagnetic layer to non-magnetic metallic layer by spin pumping \cite{PhysRevB.66.224403}. The polarization of the pumped spin current 
\begin{align}
\textbf{J}_s=g_{\uparrow\downarrow} \textbf{m}\times \dot{\textbf{m}},
\end{align}
depends on the dynamics of magnetic moments at the ferromagnetic layer \cite{PhysRevLett.88.117601}. Here, $\textbf{m}$ is the normalized magnetization direction and $g_{\uparrow\downarrow}$ is a complex value with a negligible imaginary term \cite{PhysRevB.76.104409,PhysRevLett.111.176601}. 
At the magnetic interface, the interaction between magnetization dynamic and conduction spin is dominated by the exchange interaction. Because of that, the spin polarization arises from linear response to magnetization dynamic due to s-d exchange interaction \cite{PhysRevB.96.144434}. However, the exchange interaction is smaller when the magnetic layer is antiferromagnetic because of zero-averaged magnetization \cite{Gomonay2014}. 

Recently, the spin current has been experimentally observed to arise from the dynamics of nuclear magnetic moments of MnCO$_3$ at the interface of MnCO$_3$ and Pt \cite{Shiomi2018}. MnCO$_3$ is antiferromagnetic \cite{Richards1964} and its magnetization arises from the nuclear spins \cite{Svistov_1993,Shaltiel1964}. However, theoretical mechanism of spin pumping from nuclear spin is not well understood yet. The theoretical description of spin pumping only take into account the exchange interaction between localized spin and conduction spin \cite{PhysRevB.96.144434,PhysRevB.103.094420}. Because of that, a better understanding of the mechanism spin transfer is required. Furthermore, a more efficient spin energy transfer may be achieved because nuclear spin has a lower excitation energy \cite{Kikkawa2021}.

In this article we aim to give theoretical description of the spin pumping that arise from the hyperfine interaction between nuclear and conduction spins. The article is organized as follows. In the Sec.~\ref{sec2} the linear response of the spin due to the dynamics of the nuclear spin is described. The generated spin current at the interface of antiferromagnetic layer and heavy metal layer is estimated in Sec.~\ref{sec3}. Lastly, we conclude our findings in Sec.~\ref{sec4}. Understanding of the mechanism of nuclear spin pumping may lead to applications in other fields in which nuclear spins is useful, such as quantum information \cite{Hirayama_2006,Jung2021} and medical imaging\cite{Mansfield1977,Zhang2008}.

\section{Anisotropic hyperfine interaction between nuclear and conduction spin}\label{sec2}

Spin pumping was originally described using scattering theory in Ref.~\cite{PhysRevLett.88.117601} for ferromagnet$\vert$non-magnetic interface. However, when the magnetic layer is antiferromagnetic, such as the interface with MnCO$_3$, the electron is exchange-coupled to the zero averaged-magnetization and spin pumping vanishes \cite{PhysRevB.96.054436}. While modification of scattering theory for antiferromagnetic spin pumping have been suggested \cite{PhysRevLett.113.057601,PhysRevB.90.094408}, scattering theory is not well suited to describe microscopic interaction that contains anisotropy. In order to describe anisotropic spin pumping interfaces with distorted crystal field, a linear response theory of spin pumping was developed \cite{PhysRevB.96.144434}. Linear response theory focuses on the dynamic of the spin densities of the conduction electron $\textbf{s}$. 

\begin{figure}[h]
\centering
\includegraphics[width=.5\textwidth]{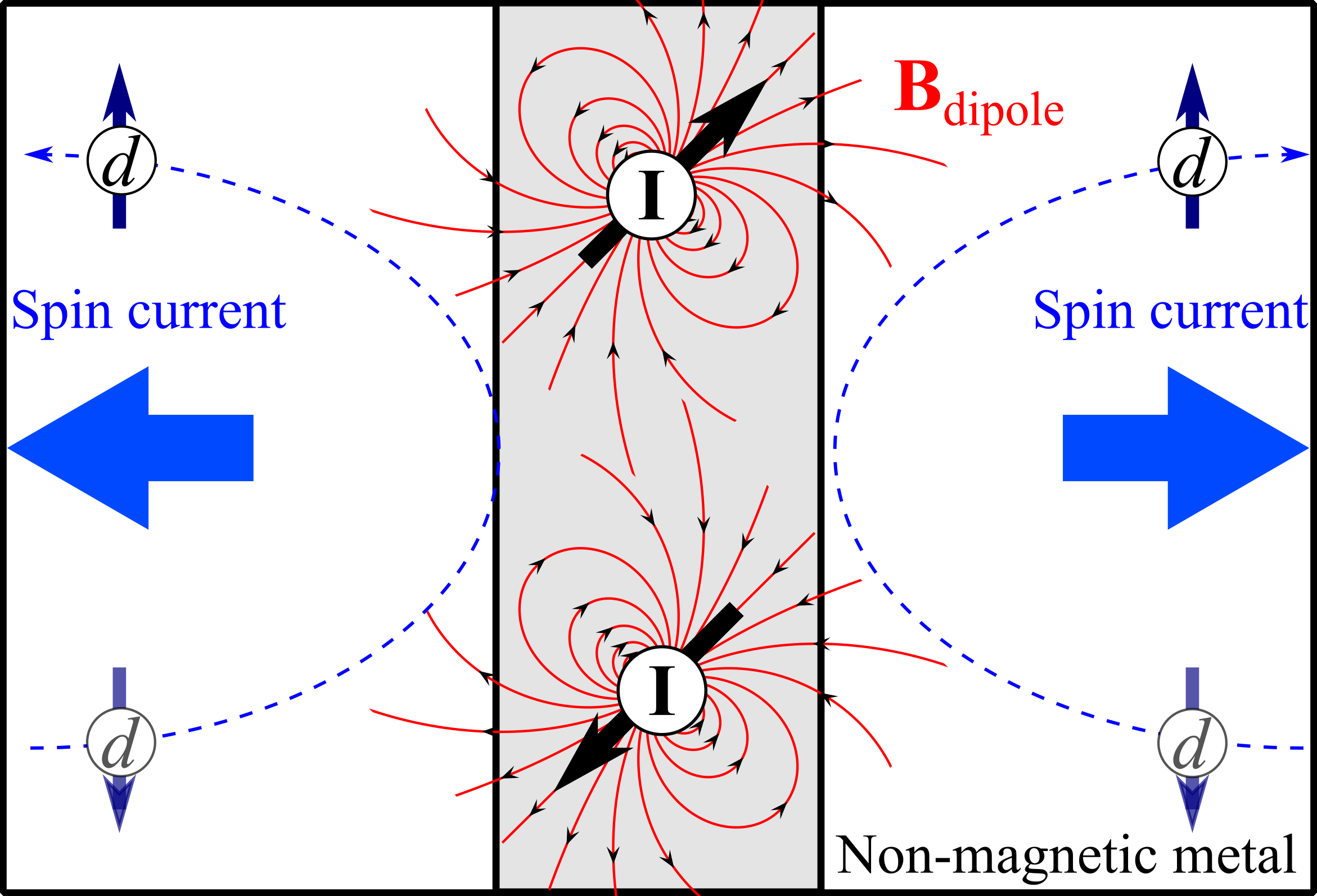}
\caption{ Microscopic picture of spin pumping by nuclear spin. A magnetic layer act as a perturbation in the metallic system.  For MnCO$_3\vert$Pt, the hyperfine interaction can be expressed by dipole magnetic field $\textbf{B}_{\rm dipole}$. The conduction electrons (d electrons of Pt) experience spin-flip reflection near the interface due to hyperfine interaction with the nuclear spin $\textbf{I}$.}
\label{Fig1}
\end{figure}

The linear response theory describes the spin polarization $\textbf{s}$ at the non-magnetic layer due to magnetic field $\textbf{B}$ from ferromagnetic layer as illustrated in Fig.~\ref{Fig1}. The linear response theory gives a simple picture of a magnetic interface. The local magnetic moment of magnetic layer and conduction spin of non-magnetic layer are interacting near the magnetic interface. The spin current generation due to this interaction can be determined from the spin angular momentum loss due to the relative direction between conduction spin $\textbf{s}$ and $\textbf{B}$
\begin{align}
\textbf{J}_s(t)=\int d^3\textbf{r}\textbf{s}(\textbf{r},t)\times \textbf{B}(\textbf{r},t)\equiv g_{\uparrow\downarrow} \textbf{m}(t)\times \dot{\textbf{m}}(t). \label{Eq.SxB}
\end{align}
When the magnetic layer is ferromagnetic, $\textbf{B}$ is dominated by effective magnetic field from s-d exchange interaction. For MnCO$_3$, the magnetic field $\textbf{B}(\textbf{r},t)$ should be appropriately modified by hyperfine magnetic field from nuclear spin.

$\textbf{B}(\textbf{r},t)$ arise from hyperfine interaction with the spin of an atomic nucleus. The hyperfine interaction can be written in the following magnetic hyperfine Hamiltonian ~\cite{book:1567088}
\begin{align}
H=&-\gamma_s\int d^3\textbf{r}\textbf{s}(\textbf{r},t)\cdot \left(\textbf{B}_{0}(\textbf{r},t)+\textbf{B}_{\rm dipole}(\textbf{r},t) \right),\notag\\
\textbf{B}_{0}(\textbf{r},t)=&\mu_0\gamma_I\frac{8\pi}{3}\textbf{I}(t)\delta(\textbf{r}),\notag\\
\textbf{B}_{\rm dipole}(\textbf{r},t)=&\mu_0\gamma_I\frac{3\hat{\textbf{r}}\left(\textbf{I}(t)\cdot \hat{\textbf{r}}\right)-\textbf{I}(t)}{r^3},
\end{align}
where $\gamma_s\textbf{s}$ and $\gamma_I \textbf{I}$ are the conduction spin and nuclear spin magnetic moments, respectively. The delta function term is the Fermi contact term that are not zero only when the conduction electron wave function can reach the nucleus \cite{Kutzelnigg1988}. Such case can be achieved when the conduction electron is from s-orbital electrons. However, in Pt, which is often used in spin pumping experiment, the conduction electron is dominated by d-electrons \cite{Sitorus2021}. Therefore, the interaction is dominated by the dipole-dipole interaction between the conduction spin from d electron of Pt and the nuclear spin $I$ of MnCO$_3$. Therefore, magnetic field in Eq.~\ref{Eq.SxB} is the dipole magnetic field $\textbf{B}_{\rm dipole}$.
The expression of $\textbf{s}$ in Eq.~\ref{Eq.SxB} can be obtained using linear response relation between the conduction spin and magnetic field. According to RKKY theory, $\textbf{s}$ is determined by a susceptibility function $\chi(r)$ \cite{RudermanKittel, Kasuya, Yosida}
\begin{align}
\gamma_s\textbf{s}(\textbf{r})= \int d^3\textbf{r}'dt' \chi (\textbf{r}-\textbf{r}',t-t') \textbf{B}(\textbf{r}',t').\label{Eq.linearresponse}
\end{align}
Here we note that in the original RKKY interaction, the magnetic field arises from s-d exchange interaction. In our case, the conduction electron is spin-polarized by dipole-dipole interaction. 

While the polarization by static magnetizations in magnetic multilayer structured has been studied \cite{Bruno1993,PhysRevB.36.3948}, it needs to be modified to describe a time-dependent magnetization. In Ref.~\cite{PhysRevB.96.144434}, the RKKY approach is generalized to ferromagnetic sheet with variable magnetization perturbing the normal metal system. By including a frequency-dependent adiabatic correction of the scalar magnetic susceptibility \cite{PhysRevB.96.144434}
\begin{align}
\chi(r,\omega)=&\chi_r(r)+i\omega\chi_i(r)\notag\\
=& \frac{N_e}{2\pi r^3} \left(\frac{\sin 2k_Fr}{2k_Fr}-\cos 2k_Fr\right)
+i\omega \frac{N_e^2\hbar\pi}{2} \frac{\sin^2k_Fr}{k_F^2r^2}, \label{Eq.chiRI}
\end{align}
it is found that there is an additional term in the spin density due to the temporal derivative of magnetic field  
\begin{align}
\gamma_s\textbf{s}(\textbf{r},t)=& \int d^3\textbf{r}' \chi_r (\textbf{r}-\textbf{r}') \textbf{B}(\textbf{r}',t)
-\int d^3\textbf{r}' \chi_i (\textbf{r}-\textbf{r}') \dot{\textbf{B}}(\textbf{r}',t) \label{Eq.adiabatic}.
\end{align}
Here $k_F$ is Fermi wave vector of Pt conduction electron and $N_e$ is electronic density of state at Fermi level.

For convenience, we can utilize the following expression of $\textbf{s}$ in momentum space
\begin{align}
\gamma_s\textbf{s}(\textbf{k},t)= \chi_r (\textbf{k}) \textbf{B}_{\rm dipole}(\textbf{k},t)-\chi_r (\textbf{k}) \dot{\textbf{B}}_{\rm dipole}(\textbf{k},t) \label{Eq.8},
\end{align}
where 
\begin{align}
\chi_r(k)=& N_e \left(1+\frac{4k_F^2-k^2}{4k_Fk}\ln \left\vert \frac{k+2k_F}{k-2k_F} \right\vert\right),\notag\\
\chi_i(k)=& \frac{N_e^2\hbar\pi^3}{2k_F^2} \frac{\Theta(2k_F-k)}{k}, \label{Eq.10} 
\end{align}
and $\Theta(x)$ is Heaviside step function \cite{Kim1999}. The Fourier transform of $\textbf{B}_{\rm dipole}$ has been widely studied because of the usefulness of dipole-dipole energy  \cite{Jiang2014,Carles2008,Lahaye2009,PhysRevA.63.053607}
\begin{align}
\textbf{B}_{\rm dipole}(\textbf{k},t)=& \mu_0\gamma_I\frac{4\pi}{3}\left(3\hat{\textbf{k}}\left(\textbf{I}(t)\cdot \hat{\textbf{k}}\right)-\textbf{I}(t)\right)\label{Eq.9}.
\end{align}
Since spin pumping only happen when the vector product of $\textbf{s}$ and $\textbf{B}$ is not zero (see Eq.~\ref{Eq.SxB}), we can focus on $\chi_i$ term
\begin{align}
\gamma_s\textbf{s}(\textbf{r},t)= -\mu_0\gamma_I\frac{4\pi}{3}\int \frac{d^3\textbf{k}e^{i\textbf{k}\cdot\textbf{r}}}{(2\pi)^3}\chi_i(k) \left(3\hat{\textbf{k}}\left(\dot{\textbf{I}}(t)\cdot \hat{\textbf{k}}\right)-\dot{\textbf{I}}(t)\right) +\mathcal{O}(\textbf{I}). \label{Eq.11} 
\end{align}
This integral can be evaluated using the following plane wave expansion \cite{Cahaya2021}
\begin{align}
e^{i\textbf{k}\cdot\textbf{r}}=4\pi \sum_{l=0}^\infty i^lj_l(kr)\sum_{m=-l}^l Y^*_{lm}(\theta_k,\varphi_k) Y_{lm}(\theta,\varphi),
\end{align}
where $Y_{lm}$ is the spherical Harmonics functions.

\begin{figure}[b]
\centering
\includegraphics[width=8cm]{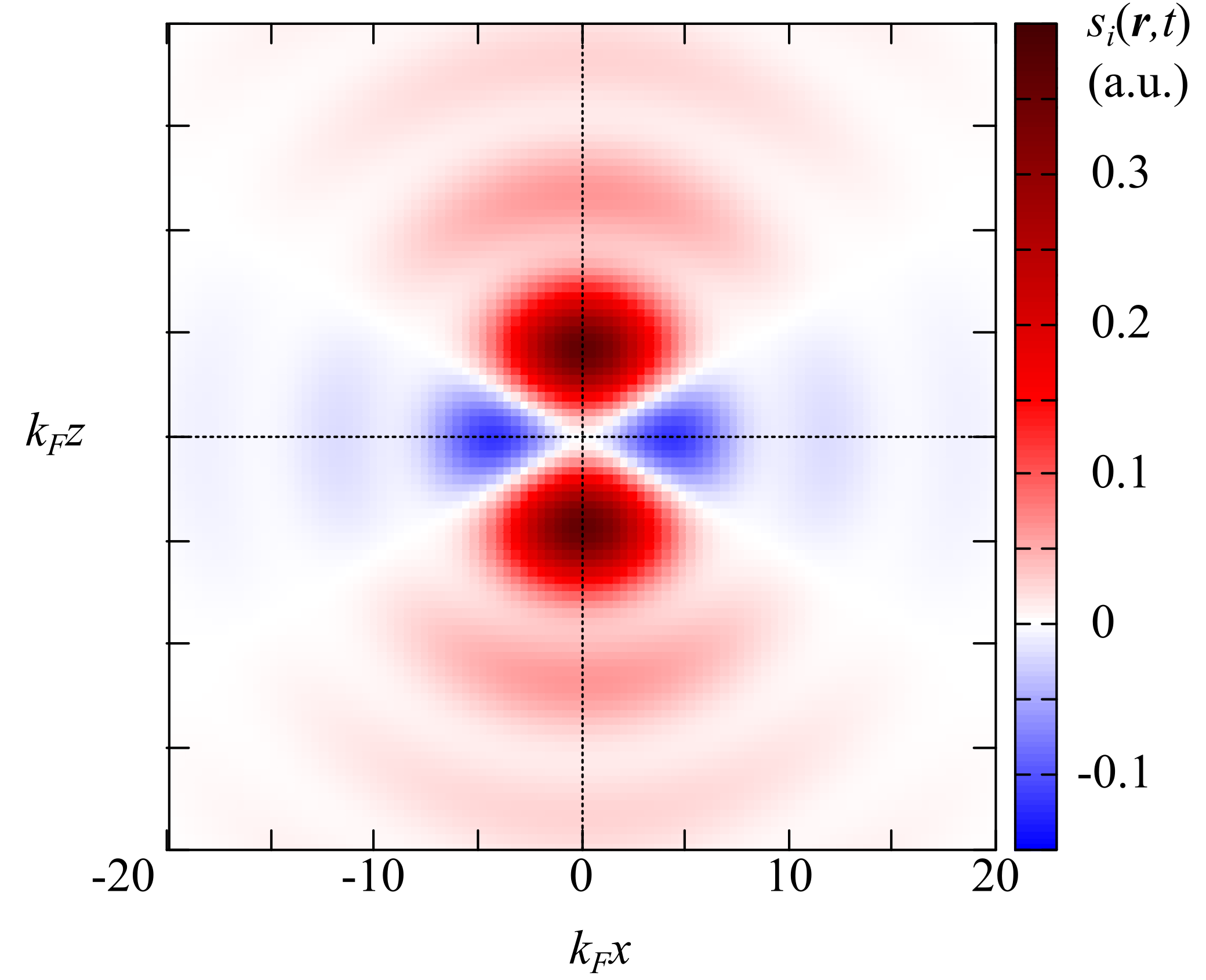}
\caption{ Spin polarization of the conduction electron spin density $\textbf{s}(r)\propto -\dot{\textbf{I}} (3\cos^2\theta-1)\left(2k_Fr(2+\cos 2k_Fr)-3\sin 2k_Fr \right)/(2k_Fr)^3$ induced by a time derivative of nuclear spin $\dot{I}$ (which is set to be in $z$-direction). The conduction electron has quadrupole anisotropy and $2k_Fr$ RKKY-oscillation, where $k_F$ is Fermi wavevector. }
\label{Fig2}
\end{figure}

The resultant $\textbf{s}(\textbf{r},t)$ has a quadrupole anisotropic and $2k_Fr$ RKKY-oscillation, as illustrated in Fig.~\ref{Fig2}  
\begin{align}
\textbf{s}(\textbf{r},t)= -\dot{\textbf{I}}(t) \frac{4\pi^2 \mu_0\gamma_I N_e\hbar}{3} \frac{3\cos^2\theta_{\dot{I}} -1}{(2k_Fr)^2} \left(2+\cos 2k_Fr -\frac{3\sin 2k_Fr}{2k_Fr} \right) \label{Eq.12}, 
\end{align}
where $\theta_{\dot{I}}$ is the angle between $\textbf{r}$ and $\dot{\textbf{I}}$. We note here that quadrupole anisotropy also occurs when the ferromagnetic layer is based on transition metal \cite{PhysRevB.96.144434} or rare earth magnetic moments \cite{voltage} due to the quadrupole anisotropy of d or f electron clouds. However, the anisotropy direction of transition metal and rare earth based spin pumping are coupled to the crystal field anisotropy direction and orbital moment direction, respectively \cite{500001078235}. Furthermore, the hyperfine-interaction-induced quadrupole anisotropy (of the neglected $\textbf{I}$ term) may contribute to the observed large magnetic anisotropy of manganese alloys \cite{Okabayashi2020}.

\section{Nuclear spin pumping}\label{sec3}

The spin pumping from a single nuclear spin can be determined by substituting Eq.~\ref{Eq.12} to Eq.~\ref{Eq.SxB}
\begin{align}
\textbf{J}_s(t)
=& -\mu_0\gamma_I\int \frac{d^3\textbf{k}}{(2\pi)^3}\chi_i(k) \frac{4\pi}{3}\left(3\hat{\textbf{k}}\left(\dot{\textbf{I}}(t)\cdot \hat{\textbf{k}}\right)-\dot{\textbf{I}}(t)\right)\times\int d^3\textbf{r} e^{i\textbf{k}\cdot\textbf{r}} \textbf{B}_{\rm dipole}(\textbf{r},t)\notag\\
=& -\frac{\left(4\pi \mu_0\gamma_I\right)^2}{9}\int \frac{d^3\textbf{k}}{(2\pi)^3} \chi_i(k) \left(3\hat{\textbf{k}}\left(\dot{\textbf{I}}(t)\cdot \hat{\textbf{k}}\right)-\dot{\textbf{I}}(t)\right)\times \left(3\hat{\textbf{k}}\left(\textbf{I}(t)\cdot \hat{\textbf{k}}\right)-\textbf{I}(t)\right). \label{Eq.orthogonalanisotropic}
\end{align}
Due to the orthogonality of two terms with quadrupole anisotropies, the integration over $\textbf{k}$ gives an isotropic $\textbf{J}_s$.  
For MnCO$_3\vert$Pt interface, parameters related to nuclear spin ($\gamma_I$ and $I$) are properties of MnCO$_3$ layer. Meanwhile, $\chi_i$ is the properties of conduction spin of Pt layer.
The discussion so far only counts for spin current generation by a single nuclear spin. In a spin pumping, the spin current that is pumped into non-magnetic metal arise from arrays of magnetic moments near the interface, as illustrated in Fig.~\ref{Fig1}.

To count for the contribution of all nuclear spins at the interface, we can assume that the nuclear spins are distributed periodically at the interface at $\textbf{r}_n$, with  $n=1,\cdots N$. Eq.~\ref{Eq.linearresponse} should be modified to take into account of all the nuclear spins at the interface
\begin{align}
\gamma_s\textbf{s}(\textbf{r},t)=& \sum_{n=1}^N\int d^3\textbf{r}'dt'\chi (\textbf{r}-\textbf{r}',t-t') \textbf{B}_{n-\rm dipole}(\textbf{r}'-\textbf{r}_n,t'),\notag\\
\textbf{B}_{n-\rm dipole}(\textbf{r},t)=&\mu_0\gamma_I\frac{3\hat{\textbf{r}}\left(\textbf{I}_n(t)\cdot \hat{\textbf{r}}\right)-\textbf{I}_n(t)}{r^3}.
\end{align}
Using this expression, the total spin pumping can be determined
\begin{align}
\textbf{J}_s(t)
=& -\frac{\left(4\pi \mu_0\gamma_I\right)^2}{9\gamma_s}\sum_{n=1}^N\sum_{n=1}^Ne^{i\textbf{k}\cdot(\textbf{r}_n-r_m)}\int \frac{d^3\textbf{k}}{(2\pi)^3} \chi_i(k) \left(3\hat{\textbf{k}}\left(\dot{\textbf{I}}_n(t)\cdot \hat{\textbf{k}}\right)-\dot{\textbf{I}}_n(t)\right)\notag\\
&\times \left(3\hat{\textbf{k}}\left(\textbf{I}_n(t)\cdot \hat{\textbf{k}}\right)-\textbf{I}_n(t)\right).
\end{align}
Since the antiferromagnetic coupling dictates that $(-\textbf{I}_n)\times(-\dot{\textbf{I}}_n)=\textbf{I}_n\times\dot{\textbf{I}}_n$, the spin current from each antiferromagnetic moments do not cancel out. 
Furthermore, due to the localization of $\chi$
\[\sum_m\chi(\textbf{r}-\textbf{r}_n+\textbf{r}_m)\simeq \chi(\textbf{r}-\textbf{r}_n),\]
we can assume that only nearest $n$ contributes to $\textbf{s}$ and the double summation reduce to the total number of the nuclear spins.
Therefore, the total spin current per unit area $A$ only depends on the density of nuclear spin at the interface
\begin{align}
\frac{\textbf{J}_s(t)}{A}
=& -\frac{N}{A}\frac{\left(4\pi\mu_0 \gamma_I\right)^2}{9\gamma_s}\int \frac{d^3\textbf{k}}{(2\pi)^3} \chi_i(k) \left(3\hat{\textbf{k}}\left(\dot{\textbf{I}}(t)\cdot \hat{\textbf{k}}\right)-\dot{\textbf{I}}(t)\right)\times \left(3\hat{\textbf{k}}\left(\textbf{I}(t)\cdot \hat{\textbf{k}}\right)-\textbf{I}(t)\right)
\end{align}
By evaluating the integral, we arrive at the expression similar to those of isotropic exchange interaction induced spin pumping.
\begin{align}
\textbf{J}_s(t)
=& N\frac{32\left(\mu_0\gamma_IN_e\right)^2\hbar}{9}\textbf{I}(t)\times \dot{\textbf{I}}(t).
\end{align}
For MnCO$_3\vert$Pt interface, $I=5/2$, $N\sim 0.05$ \AA$^{-2}$, $\gamma_I/\gamma_s\sim10^{-3}$ \cite{Shiomi2018} $N_e\sim 10^{30}$m$^{-3}$.
The spin mixing conductance of the nuclear spin pumping can be estimated to be comparable to those that arise from the exchange interaction at Y$_3$Fe$_{5}$O$_{12}\vert$Pt ($\sim10^{18} \textrm{ m}^{-2}$).
\begin{align}
g_{\uparrow\downarrow}=N\frac{32\left(\mu_0 \gamma_IN_e\right)^2 I(I+1)}{9}\sim 10^{17} \textrm{ m}^{-2}.
\end{align}
In spin pumping via hyperfine interaction, $g_{\uparrow\downarrow}$ represents the rate of spin-flipped reflection of conduction electron near the magnetic interface. Although the conduction spin has quadrupole anisotropy, due to orthogonality of the terms inside the integral in Eq.~\ref{Eq.orthogonalanisotropic}, the nuclear spin pumping is isotropic, similar to those of conventional exchange spin pumping, and its magnitude that depends on the hyperfine interaction strength, as observed in Ref.~\cite{Shiomi2018}.

\section{Conclusion}\label{sec4}

To summarize, we study the mechanism of spin pumping in bilayer MnCO$_3\vert$Pt. The spin pumping arises from the hyperfine interaction between the nuclear spins of Mn and conduction electron of Pt. The hyperfine interaction is dominated by dipole-dipole interaction. By using linear response theory, we show that dipole-dipole interaction induces a quadrupole anisotropy on the conduction electron spin. 

The spin pumping from MnCO$_3$ layer to Pt layer are governed by the nuclear spins near the interface. Due to the localization of the nuclear spins and susceptibility $\chi$, the resultant spin current only depends on the density of exposed nuclear spins. Furthermore, due to $\textbf{I}_n\times\dot{\textbf{I}}_n$ symmetry, the spin pumping of antiferromagnetic coupled nuclear spins, such as in MnCO$_3$ do not cancel out.

\backmatter



\section*{Declarations}

\begin{itemize}
\item This work was supported by FMIPA UI Grant No. NKB-001/UN2.F3/HKP.05.00/2021
\item The author have no relevant financial or non-financial interests to disclose.
\item All data is provided in full in the results section of this paper.
\end{itemize}

\begin{appendices}

\end{appendices}


\bibliography{ref}

\end{document}